# Bioinspired interfacial materials with enhanced drop mobility: From fundamentals to multifunctional applications

*Chonglei Hao, Yahua Liu, Xuemei Chen, Jing Li, Mei Zhang, Yanhua Zhao, and Zuankai Wang\**

Department of Mechanical and Biomedical Engineering, City University of Hong Kong, Hong Kong

Keywords: bioinspired interfacial materials, drop bouncing/jumping, superhydrophobic, condensation, anti-icing

**Abstract**

The development of bio-inspired interfacial materials with enhanced drop mobility that mimic the innate functionalities of nature will have significant impact on the energy, environment and global healthcare. In spite of extensive progress, the state of the art of interfacial materials have not reached the level of maturity sufficient for industrial applications in terms of scalability, stability and reliability, which are complicated by their operating environments and lack of facile approaches to exquisitely control the local structural texture and chemical composition at multiple length scales. In this review, we focus on the recent advances in the fundamental understanding as well as practical applications of bio-inspired interfacial materials, with an emphasis on the drop impact induced bouncing and coalescence induced jumping behaviors. We also suggest our own perspectives on how to catalyze new discoveries and to foster technological adoption to move this exciting area forward.

## 1. Introduction

Bio-inspired engineering, an exciting multidisciplinary research area via the transfer of function from the biological inspiration, has historically transformed the way we live and will continue to lead the way for innovation across all the aspects of science and technology. Indeed, millions of years of evolution have endowed the biological world an open-access, friendly and powerful laboratory for human to learn and imitate. By borrowing clues from various animals and plants in nature, researchers and engineers are developing artificial materials with new level of functionalities, ranging from the light-weight tough materials,[1, 2] super adhesive,[3, 4] and optical imaging[5-7] to biomedical tissues.[8, 9] In particular, at the heart of bio-inspired engineering, bio-inspired interfacial materials with enhanced drop mobility properties are of importance for a wide spectrum of industrial applications such as energy savings, green environment and healthcare.[10, 11] Thanks to the exciting progress in high speed imaging and nanotechnology,[12, 13] this exciting area has seen a new revival over the past decades. It is remarkable to note that 3 out of top 10 sleeping beauties in science, defined as publication exhibiting a long hibernation period followed by a sudden spike of popularity, are related with bio-inspired interfacial materials.[14] Currently, interfacial materials with enhanced liquid repellency properties can be made from any classes of materials, including silicon, polymer, and metal and so on. In spite of progress and vigor, ironically, these artificial materials have not reached the level of maturity in terms of scalability, stability and reliability and as a result, their applications in industrial settings remain limited, partially due to the complexity involved in the operating environment. Thus, the development of robust materials with intricate transport characteristics (hydrodynamic or thermofluidic) and the fundamental understanding of the associated physical phenomena are vital to generate significant scientific, economic, and societal impact.

In this review, we will focus our attention on the recent progress in the engineering of artificial surfaces with enhanced drop mobility inspired by lotus leaves effect[15, 16] and pitcher plant effect.[17, 18] We will review the fundamental physics underlying the drop dynamic interactions with these interfaces, including drop spreading, retraction and bouncing as well as drop nucleation, coalescence and jumping. We also describe how to promote drop mobility by controlling the surfaces roughness, wettability and ambient conditions. Furthermore, recent progress in the exploration of various bioinspired interfacial materials for condensation and anti-icing applications will be discussed. Finally, we briefly summarize our personal views for the future development within this field of research.

## 2. Fundamentals of drop mobility on bioinspired materials

Many biological systems in nature orchestrate a high level of functionalities and adaptability to their environments enabled by the exquisite manipulation of liquid states at their interfaces. Superhydrophobic surfaces (SHS)[19] with water static contact angle (CA) larger than 150°, as exemplified by the lotus leave (**Figure 1**a), allow the liquid drop to leave the surfaces easily. The extremely high drop mobility is ascribed to the combination of topographically hierarchical structure (Figure 1b) and low surface energy.[15, 20] Some animals in nature also evolve with intriguing liquid repellent surfaces, as evidenced by water striders (Figure 1c) capable of water walking/jumping induced by their non-wetting legs composed by oriented, microscale setae (Figure 1d),[21] butterfly wings allowing for directionally shedding of liquid drops,[22] and mosquito eyes with the ability to keep dry in moisture condensation circumstance[23] etc. In addition to the lotus leaves-like SHS, nature also takes advantages of another conceptually different means to enable liquid repellency. *Nepenthes* pitcher plant (Figure 1e) has evolved with the ability to trap, capture and digest arthropods aided by its unique smooth, stable liquid interfaces (Figure 1f).[17] This kind of surface also eliminates the contact line pinning of various liquids, avoids the pressure-induced impalement problem encountered in the SHS.

Over the past decade, there has seen a significant explosion in both the fundamental understanding of drop dynamics on natural and bioinspired structured surfaces and the translation of insights for practical applications (Figure 1g-i).[24-26] More relevance with practical application is the dynamic behaviors of liquid drop on these structured surfaces. For example, drop impingement is widely encountered in the airplane, pathogen transmission,[27, 28] environmental aerosol dispersion,[29] blood pattern analysis,[30] pesticides deposition,[31] and electronic spray cooling, and drop jumping is ubiquitous in the condensation heat transfer systems.[32] Thus, the understanding of the transient, multiscale processes in the context of their applications is scientifically and technologically important.

**2.1 Drop bouncing**
Drop striking a dry solid surface manifests various outcomes, depending on the surface roughness, wettability, temperature of the substrate as well as the ambient conditions.[33-37] The interaction between the liquid and underlying surfaces is a transient, dynamic process involving multiple time and length scales. Strictly, it is the local liquid-solid interaction that gives rise to the collective behavior. Here, without moving into the local hydrodynamics and non-hydrodynamic interaction between the liquid and solid, we place emphasis on the global drop dynamics. **Figure 2** summarizes recent work[38-53] on the drop impingement based on two dimensionless parameters: the Weber number $We = \rho v^2 D_0/\gamma$, which measures the relative importance of the fluid's inertia compared to its surface tension, and the Reynolds number $Re = \rho v D_0/\mu$, which measures the relative importance of the fluid inertia compared to its viscosity. Here, $\rho$, $D_0$, and $v$ represent the density, diameter and impact velocity of the liquid drop, respectively, and $\gamma$ and $\mu$ are the surface tension and dynamic viscosity of the liquid drop, respectively. At higher $We$ and $Re$, the drop disintegrates into two or more secondary drops or drop splash occurs. At lower $We$ region, the drop exhibits a quasi-elastic behavior characterized by small drop deformation and a large restitution coefficient due to negligible energy dissipation.[42] For the drop hitting solid surfaces with medium range of $We$, it is subject to a pronounced deformation driven by a pressure gradient within the time scale of $D_0/v$, leading to the redistributed liquid flow from vertical to horizontal direction.[54] In the spreading stage, the liquid drop develops into expanding lamella with a circular growing rim surrounded at the edge of lamella. Specifically, two regimes have been identified, with a defined impact number $P = WeRe^\alpha < 1$ corresponding to the capillary-inertial regime where the kinetic energy is mainly transformed into the surface energy, and viscous-inertial regime ($P > 1$) where the initial kinetic energy is mainly dissipated by the viscous force. Note that the value of α is still under debate: it can be either -2/5 based on the energy conservation model or -4/5 based on the momentum conservation model.[38] In the viscous-inertial regime,[38, 55] the maximal spreading ratio $D_{max}/D_0$ is assumed to follow the scaling law of $Re^{1/5}$. In the capillary-inertial regime, however, $D_{max}/D_0$ follows the scaling law of $We^{1/2}$ (for α = -2/5) or $We^{1/4}$ (for α = -4/5).[38, 56, 57]

One of the most important parameters characterizing the drop dynamics is the contact time, defined as the time duration between the moment of drop impacting and bouncing off the surfaces. Despite its complexity at first sight, the contact time associated with an impinging drop on SHS can be simplified by considering its inertial-capillary nature. Briefly, the impinging drop can be treated as a free oscillation system with stiffness $\gamma$ and mass $\rho D_0^3$, and the contact time $\tau$ is scaled as the inertial time scale $\tau_i = \sqrt{\rho D_0^3/\gamma}$.[58, 59] Note that the time spent in the spreading stage only accounts for about 20 ~ 30 % of the total contact time. Different from the inertia-dominant spreading process, the capillarity serves as the driving force and the majority of the contact time is dissipated in this retraction stage. The drop retraction rate ($V_{ret}/D_{max}$) is dependent on the relative importance of viscous force to the inertial and surface tension forces, or Ohnesorge number $Oh = \mu/\sqrt{\rho D_0 \gamma}$. For the case of $Oh < 1$, based on the Taylor-Culick method,[60, 61] the retraction rate at the early stage of retraction process is calculated as

$$V_{\text{ret}}/D_{\max} \sim \sqrt{1 - \cos\theta_R}/\tau_i$$

where $\theta_R$ is the receding CA. However, for $Oh > 1$, the retraction rate follows the scaling of [62]

$$V_{\text{ret}}/D_{\max} \sim \mu D_0/\gamma .$$

One of the most important issues hindering the complete bouncing of drop from structured surfaces is the Cassie-Wenzel transition,[19, 52, 63-68] which is characterized by the breakdown of the global superhydrophobic property. Although the specific mechanism responsible for the occurrence of Cassie-Wenzel transition is still under debate, the complex picture can be illustrated by considering the pressure involved in this process. When the impinging drop is in contact with the underlying surface, it is subject to the dynamic pressure expressed as $P_D = \rho v^2/2$ and the capillary pressure $P_C$ resisting the penetration of the drop.[63, 69, 70] It is widely held that when the dynamic pressure is larger than the capillary pressure, the drop is susceptible to the Cassie-Wenzel transition. The water hammer pressure, exemplified in a pipe flow,[71, 72] was recently demonstrated to play an important role in the wetting transition.[73, 74] The water hammer pressure is produced at the contact area with a diameter estimated as $d_{\text{WH}} = D_0 v/c$ and can be expressed as $P_{\text{WH}} = k\rho vc$, where $D_0$ is the initial drop diameter, $c$ the speed of sound in the drop, and $k$ the pre-factor determined by the surface morphology, shape and velocity of the drop.[71, 74] Based on the interplay between $P_C$, $P_{\text{WH}}$ and $P_D$, drop hitting textured surfaces can display total wetting, partial wetting or the non-wetting state.[75-81] However, it is also reported by Maitra *et al.* that on textured SHS, the compressibility of draining air caused dimple formation and subsequent pressure rise between the drop and the substrate, rather than the water hammer pressure effect, should be responsible for the observed liquid meniscus penetration in the We range $10^2 \sim 10^3$.[82] Recently, Lee *et al.* observed the unexpected occurrence of Cassie-Wenzel transition at the drop retraction phase.[83] Through numerical simulation, they proposed that this new type of Cassie-Wenzel transition mode is due to the buildup of the wetting pressure ($P_s \propto \sqrt{\rho\gamma/R}v$) as a result of the vertical momentum transfer. Excitingly, structured surfaces with exquisite morphologies such as mushroom-like reentrant,[84-87] or doubly reentrant structures[88] have been demonstrated to exhibit superior water repellency and it is expected that the unwanted Cassie-Wenzel transition can be largely suppressed on those surfaces.

Note that in practical applications, most surfaces are obliquely placed and thus the study of oblique impact has received increasing attention.[89, 90] It is reported that the contact time in the case of the oblique impact is basically the same as that of the vertical impact.[91] On another note, oblique bouncing can also occur on the horizontally placed surface but with wetting or thermal gradient.[92-95]

## 2.2 Drop jumping

On another research line, the coalescence-induced drop jumping phenomenon has drawn growing attention owing to its many practical applications such as in heat transfer, anti-icing, and energy harvesting. This phenomenon has also been observed in many natural biological systems, such as the ballistospore discharge process (**Figure 3**a),[96] self-cleaning cicada wings[97] and water strider legs.[98] Drop jumping is a bottom-up process and normally occurs in the phase change process such as vapour condensation. Distinct from the drop impinging process where the drop lands on the surface from above the surface and with a size normally at the scale of ~ mm, the condensate drops grow from the bottom interstitials of the substrate with a critical nuclei of ~10 nm.[99] Thus, both Wenzel- and Cassie-state drops can be formed during the condensation process and hence SHS with excellent drop mobility do not necessarily preserve good drop jumping (Figure 3b). Without the proper control of surface wettability and structure, the SHS suffer from severe liquid flooding at the high humidity and the superior drop mobility is lost. It was previously observed that on SHS with microscale roughness alone,[100-104] condensate drops tend to nucleate and grow in the cavities between microstructures, forming sticky condensate drops in the Wenzel state. By contrast, condensate drops on SHS with proper design[105-113] could depart via coalescence-induced jumping (Figure 3c). By exquisite control over the surface roughness and wettability, drop nucleation can be spatially confined in predetermined locations, leading to well-controlled nucleation and subsequent self-removal on a rapid manner (Figure 3d).[114, 115] Accordingly, fresh nucleation sites are exposed and trigger the new cyclical process of nucleation, coalescence, and departure.

The drop jumping velocity can be scaled as $\sqrt{\gamma/\rho D_0}$ assuming all the released surface energy is converted to the kinetic energy.[116, 117] However, the measured drop jumping velocities were significantly smaller than that predicted based on the capillary-inertial scaling law. On the basis of theoretical modeling based on energy conservation of interfacial, kinetic and viscous energy, Wang *et al*. found that there exists a critical drop size of ~10 μm for the occurrence of jumping motion.[118] On the other hand, some studies reported that the surface adhesion of condensed drops is dominant in determining the critical size and the jumping velocity.[119, 120]. Numerical simulation was also conducted to provide insights for the energy estimation during the dynamic process.[121] Nam *et al*. developed a full 3D numerical model showing that about half of the released surface energy during the coalescence is converted to kinetic energy.[122] However, Enright *et al*. argued that only a small fraction of excess surface energy (6%) is transformed into translational kinetic energy.[123] Recent experimental observations reveal that the jumping of condensate drops is mainly resulting from the coalescence of multiple drops. Rykaczewski *et al*. identified several drop growth and shedding modes on the hierarchical SHS.[124] The serial coalescence events of drop culminate in formation of a drop that either departs or remains anchored to the surface. Lv *et al*. reported that multiple drops-induced coalescence is more beneficial for drop jumping owing to its relative low energy barrier.[125] The coalescent drop can also be propelled in an oblique direction aided by the release of the elastic energy stored in the deformed textures.[98] This drop jumping phenomenon also occurs on hydrophobic fibers under a proper drop-to-fiber radius ratio.[126] In addition to drop coalescence induced release of excess surface energy, the manipulation of electric field in electrowetting on dielectric (EWOD) has also been introduced to change the surface energy of liquid drop to induce the jumping motion either in ambient or oil environment.[127-129] Moreover, recent work by Schutzius *et al*. revealed a novel jumping motion of drop placed on appropriate SHS under low-pressure environment, which is caused by the overpressure underneath the drop generated by fast evaporation.[130]

**3. Approaches to enhance drop mobility**
One central point in the rational design of bio-inspired surfaces for multifunctional applications is to enable the impinging drop to depart from the surfaces as fast as possible. In other words, the contact time should be as small as possible to minimize the extent to which mass, momentum and energy are exchanged between drops and underlying surfaces. Conventional wisdom in the development of synthetic non-wetting surfaces is normally confined to surfaces with tiny structures to prevent the collapse of the air cushion. As discussed above, however, there exists a minimum contact time that was assumed to be attained for conventional SHS.[58] Over the past decade, the scientists and engineers embarked on an exciting journey to explore novel strategies to break the classical contact time. In **Figure 4**a, we summarized the milestones achieved in this area.

**3.1 Macrotexture structure for fast drop bouncing**
Bird *et al*. recently reported that the contact time can be reduced by adding macroscopic ridges on SHS to induce asymmetric and fast recoil.[131] As shown in Figure 4b, when the drop impacting on the ridges much smaller than the drop size, it splits into satellite ones and results in ~37% contact time reduction compared to that on controlled surfaces. Gauthier *et al*. and Shen *et al.* further demonstrated that by the design of macrotexture surface with different patterns such as Y-shape or cross-shape, drop can be configured into different subunits and leaves the surface with a reduced contact time (Figure 4c).[132, 133] Recently, Liu *et al* fabricated curved macrotextures (convex and/or concave) with the curvature of radius comparable to the drop size.[134] As shown in Fig. 4d, the drop spreads larger in the azimuthal direction than in the axial direction, and leaves the surface with an elongated shape along the azimuthal direction. As a result of the asymmetric momentum and mass distribution, there is a preferential fluid flows around the drop rim and the contact time was reduced by ~ 40%.[135]

In 2014, Liu *et al*. made an exciting breakthrough in developing a novel surface that achieves the counterintuitive "pancake bouncing" to reduce the contact time by 80% (Figure 4e).[136] This work challenges the conventional view of drop impact process established since one century ago. The so-called "pancake bouncing" results from the rectification of capillary energy stored in the penetrated liquid into upward motion adequate to lift the drop.[137, 138] Moreover, in the case of drop impacting on tapered posts with a diameter that increases continuously and linearly with depth in the vertical direction, the upward capillary force increases with the penetration depth which permits modeling of the capillary force as a harmonic spring. The tapered post morphology can strongly influence on the robustness of the occurrence of pancake bouncing.[139] With a larger apex angle of the tapered post arrays, the pancake bouncing can occur with smaller critical *We* and wider *We* range. Interestingly, elongated pancake bouncing also occurs on the conventional SHS in the condition of highly oblique impact.[89]

**3.2 The effect of air lubrication for fast drop bouncing**

Apart from the surface roughness, the maintaining of a stable and robust air layer which screens the contact of liquid with the solid substrate is essential to the complete bouncing of drop. In 1879, Lord Rayleigh first observed bouncing phenomenon between two colliding streams of drops,[140] which was ascribed to the maintaining of an integral air cushion.[141-145] The air cushion can be stabilized by many methods, such as the addition of surfactant to the drop,[146] the vibration of underlying liquid bath,[147] or the use of very smooth interfaces that allow for the easy air entrapment (**Figure 5**a).[148, 149] Recently, Hao *et al*. studied the impact of drop on thin liquid films and discovered a robust superhydrophobic-like bouncing, characterized by the contact time, the spreading dynamics, and the restitution coefficient independent of the underlying liquid substrate (Figure 5b).[150] The emergence of such superhydrophobic-like bouncing directly results from the sustaining a robust air layer on the soft and smooth interfaces and its small energy dissipation.[151] Interestingly, the drop bouncing on such thin liquid films[150, 152] is very similar to the work done by de Ruiter *et al*.[40] if we treat the solid substrate as a thin liquid film with viscosity close to infinity, and after the rupture of the air cushion, the drop impingement dynamics is strongly dependent on the physical nature of the liquid film, such as viscosity (Figure 5c).[153] Due to the soft nature of the substrate, a wetting ridge is formed at the triple-phase interfaces as a result of surface tension,[154-156] which can be visualized by the transmission X-ray microscopy.[157] The evolution of the air profile underneath the impinging drop can be measured based on light interferometry, such as angle shift method,[158] dual-wavelength reflection contrast microscopy (DW-RICM) imaging,[40, 148, 159-161] and white light interferometry.[162]

The air cushion entrapped between the impinging drop and the underlying substrate can be stabilized through the control of phase change process. When the substrate temperature *T* is higher than the Leidenfrost temperature $T_L$, the drop is levitated by its own evaporating vapor. Thus, dependent on the competition between the Leidenfrost temperature and substrate temperature, the impinging drop exhibits in the contact boiling regime (**Figure 6**a) or Leidenfrost regime in which the drop is fully supported by the vapor (Figure 6b-c). A phase diagram is shown in Figure 6d to summarize different behaviors of drop impingement on such a hot flat plate.[44] Note that the formation of an integral film during the Leidenfrost regime is not beneficial for evaporation based thermal management applications such as in the spray cooling of electronic devices. On surfaces with larger ratchet structures,[163-165] the vapor flow is dominant and the Leidenfrost drop shifts against the tilted direction, whereas the drop moves along the tilted direction for surfaces with tiny features (Figure 6e).[166, 167] Li *et al*.

reported that a directional rebounding towards the higher heat transfer region can be achieved through breaking the wetting symmetry at high temperature.[134] Different from the Leidenfrost effect, Antonini *et al*. recently[168] demonstrated possibility of complete bouncing originating from the solid substrate (dry ice) sublimation with temperature controlled around -79 °C, and the observed impact behaviors follow the same scaling law of those on SHS (Figure 6f).

## 4. Bioinspired interfacial surfaces with enhanced drop mobility for practical applications

The excellent drop mobility inherent with interfacial materials has sparked various applications including self-cleaning,[169-174] dropwise condensation,[175-177] anti-icing,[178-180] anti-corrosion,[181-186] drag reduction,[187-190] water/oil separation,[191-195] antimicrobial[196-200] and so on. In this review, we mainly focus on the dropwise condensation and anti-icing which is assisted by drop jumping and drop bouncing.

### 4.1 Condensation application

Condensation is a multiphase and multiscale phenomenon, and is ubiquitous in nature and many industrial applications such as desalination,[201] power generation,[202] water harvesting[203] and air conditioning system.[204] In these systems, the enhancement in the condensation heat and mass transfer could significantly improve system performance, reduce energy cost, and impact economic development. Condensation is dramatically influenced by the physical structure and chemical properties of the solid surface.[106, 205-212] Depending on the surface wettability, water vapor condenses on the solid surface exists either in the form of thin films (filmwise) or individual drops (dropwise). In filmwise condensation, the liquid film formed between the solid surface and the vapor brings an additional resistance, limiting the amount of heat transfer that can take place across the solid surface. In dropwise condensation, discrete mobile drops are formed and fall off the surface at a reduced size. It is reported that dropwise condensation can produce heat transfer coefficients that are an order of magnitude higher than in filmwise condensation.[213]

Since condensation involves a series of multiphase and multiscale processes including drop nucleation, growth, coalescence and departure, the basic philosophy in the design of enhanced condensation surfaces is to synergize these processes. In other words, an ideal condenser should (i) promote fast drop nucleation and growth, (ii) accelerate the drop removal at small length scale; and (iii) avoid or decrease the additional interfacial thermal resistance. Since these requirements are intricately intertwined together and sometimes demand conflicting requirements on the physical and chemical properties of solid surfaces, the design of efficient condenser surfaces requires a systematic method, rather than solely considering an isolated feature. This poses challenges for the design of superior condenser surfaces.

As discussed earlier, conventional SHS exhibiting excellent drop mobility are not necessarily suitable for condensation applications.[214] Although SHS enable the formation of condensate drops in the mobile Cassie state, they suffer from reduced drop nucleation density due to the high nucleation energy barrier as compared to the hydrophilic surface.[99] To reconcile these two conflicting features into one surface, Chen *et al*. fabricated a nanograssed micro-pyramidal surface (**Figure 7**a) with a hybrid wettability.[114] The sidewalls of the pyramids surface were spatially controlled to be wettable for drop nucleation and the surface is globally superhydrophobic for enhanced drop self-removal. As a result of the synergistic effect, there is ~ 65% increase in the drop number density and ~ 450% increase in the drop self-removal volume compared to the SHS with nanostructures alone. Inspired by the Namib desert beetle (*Stenocara*),[176] which utilizes its alternating hydrophilic bumps and hydrophobic troughs to condense and collect water drops, Hou *et al*. developed a hybrid surface with hydrophilic patches on the top of the micropillars which are surrounded by superhydrophobic nanograss (Figure 7b) to enhance drop nucleation as well as self-removal.[115] This hybrid surfaces lead to ~63% increase in heat transfer coefficient as compared to the homogenous hydrophobic surfaces. To increase the mass transfer, surfaces with periodically designed hydrophilic and hydrophobic patches were also reported to remove condensate drops efficiently.[215, 216] It is also notable that there are still many debates about whether the real *Stenocara* beetle actually collects fog water with fog basking behavior in nature and whether the real dorsal surface of *Stenocara* beetle is wetting contrast.[217, 218] Apart from the spontaneous drop jumping on SHS by the control of surface structure,[219] drop jumping frequency can be enhanced by application of external electric field (Figure 7c).[220, 221] Miljkovic *et al*. demonstrated the measured condensation heat transfer coefficient was increased by 50% as a result of fast drop collection compared to the surface without electric field.[220] Recently, Rykaczewski *et al*. reported the dropwise condensation on the flat copper surface embedded with hydrophobic nanoparticles patches. Owing to the flat nature of the surface, this design might avoid the liquid flooding problem encountered in the conventional SHS.[208]

Different from the condensation enhancement on SHS assisted by the spontaneous drop jumping, slippery liquid-infused porous surfaces (SLIPS) with low CA hysteresis promotes the condensation efficiency through the easy shedding of condensate drops.[18, 222-224] As shown in Figure 7d, the slippery property of SLIPS renders the condensate drops highly mobile even at the small length scale ~ 100 μm,[222] and the prompt drop removal from SLIPS refurbishes the nucleation sites in a continuous manner and increases the entire phase change processes.[224] Additionally, the SLIPS have the potential to avoid the liquid flooding problem encountered in the SHS-based condensers. However, the introduction of lubricants in SLIPS also brings additional thermal resistance and their stability for long term operation is still a challenge. Recently, organogel[225-227] or ionic-liquid-gel[228] based slippery surfaces without the need for structured matrix were also nicely developed.

**4.2 Anti-icing application**
The ability to resist icing formation is of paramount importance in practical application areas such as aircraft, transmission line, and heat transfer system. Conventional passive approaches for anti-icing/deicing involve various surface treatment processes including thermal control to improve the temperature above the nucleation point,[229] addition of chemicals to reduce the freezing point,[230] surface coating to reduce the adhesion force after the ice/frost formation,[231] and mechanical force[232] to remove the ice directly etc. Despite effective, these mentioned methods are costly and not environmentally friendly. The enhanced drop mobility associated with bio-inspired surfaces has spurred the exploitation of their anti-icing application. Technically, the design of anti-icing surfaces is more challenging than the condenser surfaces. Currently, the development of anti-icing surfaces is mainly based on the SHS or the SLIPS (organic or aqueous), with the former mainly leveraging on the advantage of trapped air cushion that minimizes the interaction between impinging drops and surfaces whereas the latter taking advantage of the reduced adhesion of ice.[233, 234]

Owing to the manifestation of enhanced drop bouncing and the existence of air cushion in their rough structure, SHS have been heralded for anti-icing applications.[235-239] Indeed, many studies have shown that SHS can repel impinging water drops and maintain the ice-free state at low temperatures down to -25 ~ -35 ˚C (**Figure 8**a-b).[178] Moreover, SHS can not only inhibit the ice nucleation via self-removal of jumping condensates as indicated in Figure 8c with the vanish of outlined cluster of drops between 141.2 and 142.2 s , but also suppress the frost spreading by reducing the interdrop ice bridge formation.[234] Recently, Chen *et al*. proposed that freezing wave propagation can be significantly suppressed via activation of the microscale edge effect in the hierarchical SHS, due to that the energy barrier for ice bridging as well as engendering the liquid lubrication during the defrosting process are increased.[240] Different from the impinging at the ambient conditions, however, the impalement of liquid drop on SHS under supercooled conditions (either drop or substrate) would greatly magnify the viscous dissipation, which can lead to unwanted increase in the contact time and the possibility of drop freezing.[82, 241] At some extreme condition such as high supersaturation, the SHS is also susceptible to the formation of conformal frost coating,[242, 243] as shown in Figure 8d. Notably, the drop freezing is closely dependent on the interplay between the critical nucleus radius dictated by the thermodynamics and the feature size of the solid surface. Jung *et al*. demonstrated that hydrophilic surfaces with nanoscale roughness close to the critical nucleus radius even exhibited higher icephobicity than typical hierarchical SHS.[244] When the feature size of surface textures is larger than the critical nucleus radius, the probability of icing is increased with increasing the feature size of surface textures.[179] It was recently reported that external environmental conditions, such as gas flow and humidity, can also significantly affect the nucleation mode (homogenous or heterogeneous) of the drop.[245] These results indicate that the wettability and surface roughness should be optimized in the context of the experimental conditions to achieve superior anti-icing behavior.

On another research line, owing to the extremely small CA hysteresis and smooth liquid interfaces, the organic lubricant in SLIPS is capable of suppressing ice/frost accretion by the timely removal of the condensate or impinging drops (Figure 8e).[180] Moreover, upon the formation of ice crystal, the ice adhesion on the SLIPS is an order of magnitude lower than those ordinary flat surfaces materials, which is beneficial for deicing/defrost process at higher temperature.[246, 247] By introducing an aqueous lubricant layer infusing the hygroscopic polymer, Chen *et al*. found that this organic-inorganic composite coating is capable of not only reducing the ice adhesion, but also depressing the freezing/melting temperature of water significantly.[248]

## 5. Future Perspectives and Conclusions

In this review, we take the integrative approach to bridge two interesting drop dynamics (impact induced bouncing and coalescence induced jumping) on the bio-inspired interfacial materials and discuss the principles, challenges underlying the materials design and recent applications propelled by the enhanced drop mobility.

Due to the space limit, it is hard for us to make a clear and cohesive description of this important topic while without bypassing other interesting topics such as omniphobic materials and many other applications (anti-dew, anti-biofouling, and water harvesting etc.). We also apologize to those authors whose work is not duly cited in our review.

Thanks to the exciting advances in high speed imaging and micro/nanotechnology, many new phenomena emerge like the grooming bamboo after the rain,[18, 41, 130, 131, 136] and the mechanisms associated with the phenomena are also being elucidated. In spite of significant advances, our understanding of wetting statics and dynamics still remains elusive. For example, as discussed, the use of macrotextured surfaces can significantly modify the hydrodynamics. However, how to bridge the global behaviors such as contact time, restitution coefficient to the local hydrodynamics including the moving contact line, the air cushion at the triple-phase and liquid lamella evolution remains to be answered. From the mass transfer and energy point of view, how do these macrostructures mediate the mass/momentum transfer and energy transformation between kinetics, elastic energy, and viscous energy? The fundamental understanding is further complicated by the phase change process. Thus, the drop mobility on the macrostructures in the context of phase change processes such as condensation and icing remains to be uncovered. Moreover, the microscale drop dynamics at both the single drop and ensemble drops levels including heterogeneous nucleation, wetting transition, coalescence of drop, jumping and freezing are not only dependent on the surface morphology and wettability, but also intricately affected by the working environment. Thus, more systematic investigations are needed to decouple the collective effects and identify the individual role. Moreover, since impurities are inevitable, the effects of these notorious particles on the local and global activities should be considered in the materials. With the development of advanced microscopy with high time and spatial resolution, it is expected that many of these fundamental questions can be answered in the years to come.

In the applied research aspect, facile and thoughtful strategies should be explored to foster the technology commercialization to push this emerging area forward. It is remarkable that many applications have been nicely demonstrated such as dropwise condensation, anti-icing, self-cleaning, anti-corrosion, drag reduction, water/oil separation, antimicrobial and so on. However, efforts should be put to the development of novel materials and fabrication processes[249] that can enable functionalities yet with the minimal barrier for large-scale technology adoption. In this sense, these technologies are indispensable, rather than only serving as the supplementary approach to the standard methods. Notably, many interfacial materials such as those proposed for dropwise condensation and anti-icing applications indeed possess unique advantages which are impossible to achieve for the conventional materials. In spite of those beneficial properties, there are still several limitations in implementing these materials in engineering settings. First, the design and fabrication of materials is complicated by the complex working environments and the diversity in their applications. This is in striking contrast to the information technology where silicon is the standard building block material. The silicon based fabrication processes are well established, and various functional devices can be integrated and work in a dry environment. By contrast, there is no standard material or surface structure suitable for all the applications in the area of interfacial materials. But, from some point of view, we can borrow the insights and lessons of emergence of microfluidics, which learned from silicon microelectronics and microelectromechanical systems (MEMS), but has gone beyond the traditional dry silicon technology to the wet conditions.[250]

So far, various surfaces with exquisite structures have been demonstrated in the laboratory; however, the real applications require excellent chemical or mechanical robustness/stability, and fabrication scalability with low cost. Thus, a wide spectrum of cheaper, accessible materials and fabrications methods are needed to translate the lab demonstrations to the market, which might favor the choice of the simple method without the use of fancy designs. Moreover, in contrast to the rich structural and functional diversity observed in nature, the state of the art of synthetic materials are limited to simple functionalities, partially due to the lack of facile approaches to exquisitely control the local structural texture and chemical composition at multiple length scales and to dynamically adapt to the external environment.[251, 252]

Considering many compelling advantages inherent in bio-inspired interfacial materials, the depth and breadth in our fundamental understanding and technology adoption as well as the closer cooperation between materials, engineers and chemists, the bio-inspired interfacial materials is embracing a bright future.

**Acknowledgements**

This work was supported by the RGC Grant (No. 11213414), the National Natural Science Foundation of China (No. 51475401, No. 51276152).

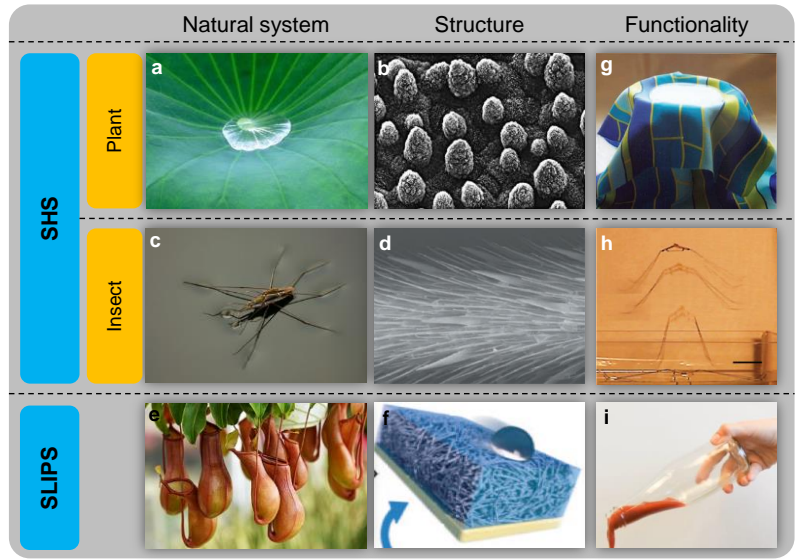

**Figure 1.** Representative natural systems with liquid repellent properties, structure and functionality. (a, b) Lotus leaves demonstrate superhydrophobic and self-cleaning properties (a) due to the presence of hierarchical micro/nano structures (b) with decoration of a layer of epicuticular waxes. Reproduced with permission.[16] Copyright 1997, Annals of Botany Company. (c, d) The water strider with the capability of water-walking (c) possesses the superhydrophobic legs composed by directional microsetae and nanogrooves (d). Reproduced with permission.[21] Copyright 2004, Nature Publishing Group. (e, f) Pitcher plant which can capture prey (e), is aided by its unique slippery liquid held in place within textures (f). Reproduced with permission.[18] Copyright 2011, Nature Publishing Group. (g-i) The multi-functional biomimetic studies of liquid-repellent surfaces for practical applications, such as the non-wetting textiles (g), water jumping robotic insects (h), and surface processing of packaged goods (i). (g) Reproduced with permission.[24] Copyright 2011, WILEY-VCH. (h) Reproduced with permission.[25] Copyright 2015, the American Association for the Advancement of Science (AAAS). (i) Image provided by LiquiGlide Inc.

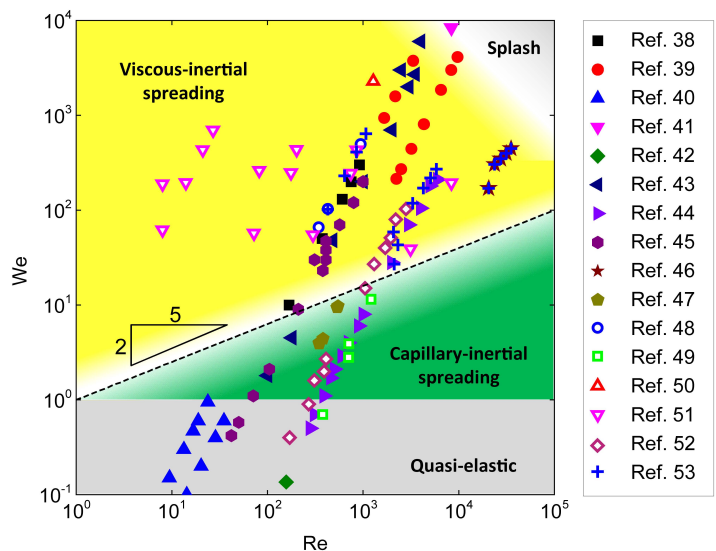

**Figure 2.** Weber and Reynolds numbers for various droplet impact experiments. The dashed line signals $P = 1$ ($P$ is defined as $P = We/Re^{2/5}$) and separates the capillary-inertial regime with $P < 1$ from the viscous-inertial regime with $P > 1$.

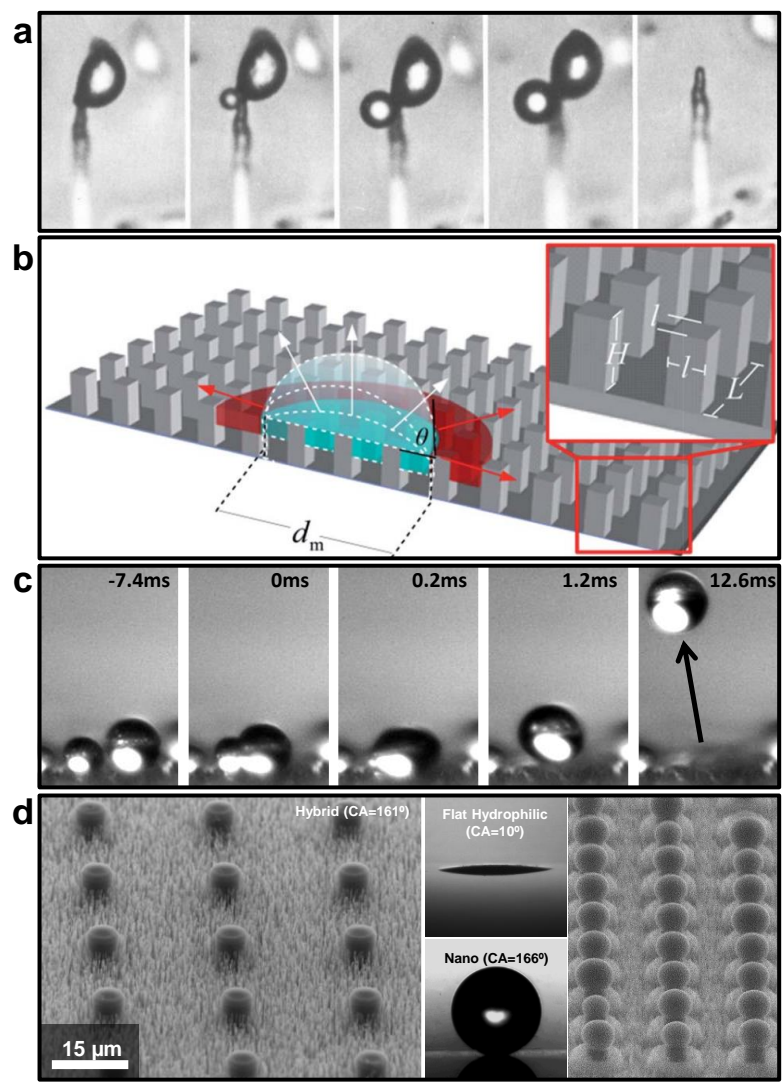

**Figure 3.** Drop jumping motion. (a) Selective images demonstrating the growth of Buller's drop at the base and liquid film on the side, and finally the drop and spore disappeared from sterigma due to the coalescence of the drop and liquid film. Reproduced with permission.[96] Copyright 1984, Elsevier. (b) Schematic diagram of a condensed drop confined within nanostructures. The drop can grow by increasing its volume in either lateral or upward direction. Reproduced with permission.[107] Copyright 2012, The Royal Society of Chemistry. (c) Side-view imaging of spontaneous drop jumping via dynamic coalescence on SHS. Reproduced with permission.[116] Copyright 2009, The American Physical Society. (d) Spatial control of droplet nucleation sites (right panel) during condensation on hybrid surfaces (left panel) composed by heterogeneously hydrophilic micropillar (CA ~ 10°) and hydrophobic fluorinated nanograss (CA ~ 166°) (middle panel). Reproduced with permission.[115] Copyright 2015, American Chemical Society.

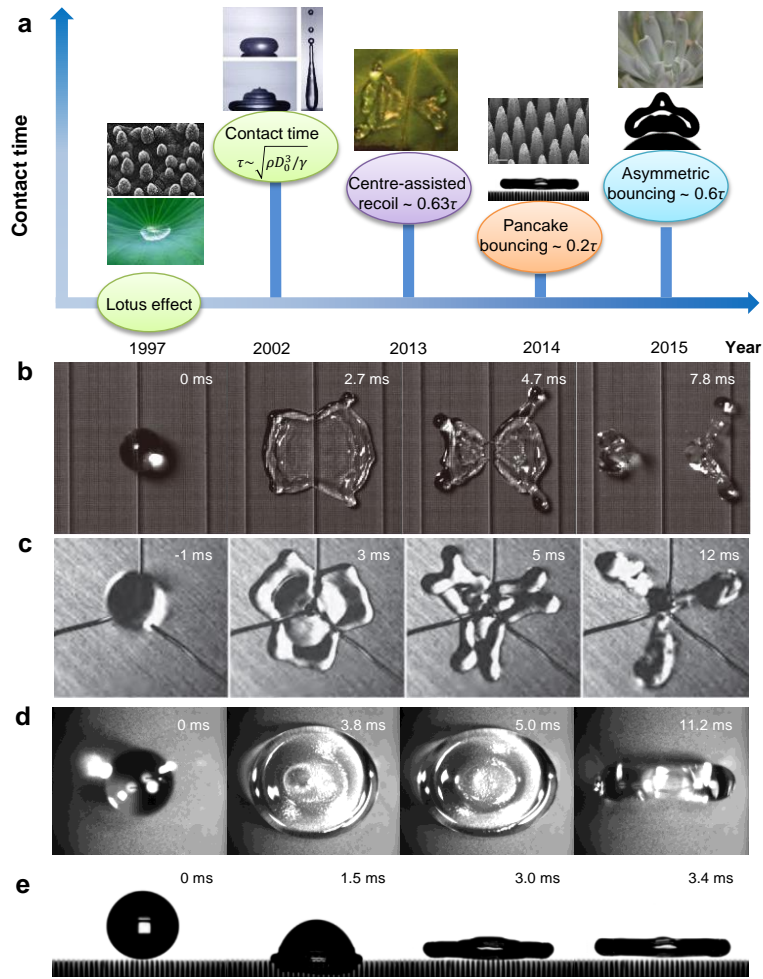

**Figure 4.** Strategies for contact time reduction. (a) The dependence of investigated contact time of drop impact on SHS on the year showing the breakthrough on how to reduce contact time by the design of novel structures. Reproduced with permission.[16, 58, 131, 134, 136] Copyright 1997, Annals of Botany Company; Copyright 2002, Nature Publishing Group; Copyright 2013, Nature Publishing Group; Copyright 2014, Nature Publishing Group; copyright 2015, Nature Publishing Group. (b) Time-lapsed images showing the contact time reduction by drop impacting on macrotextured surfaces. Reproduced with permission.[131] Copyright 2013, Nature Publishing Group. (c) Top-view images of a water drop bouncing on the superhydrophobic macrotextures with Y-pattern. Drop during impact makes six lobes that quickly merge in three symmetric subunits. Reproduced with permission.[133] Copyright 2015, Nature Publishing Group. (d) Selected snapshots showing a drop impacting on an *Echeveria* leaf with surface curvature. Reproduced with permission.[134] Copyright 2015, Nature Publishing Group. (e) Selected snapshots showing a drop impacting on the widely spaced tapered surface. The drop bounces off the surface in a pancake shape at ~3.4 ms. Reproduced with permission.[136] Copyright 2014, Nature Publishing Group.

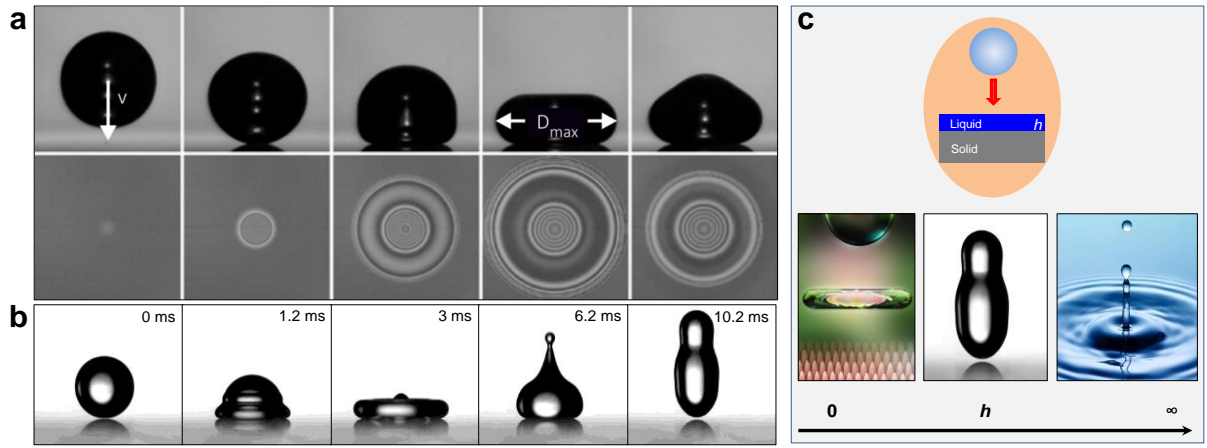

**Figure 5.** Drop bouncing mediated by the air cushion. (a) Selected images showing the drop bouncing observed on the hydrophilic surfaces with thin air cushion entrapment at We ~ 0.7 (upper row), and the synchronized RICM interference signal clearly shows the existence of air cushion during the entire drop impact process (bottom row). Reproduced with permission.[148] Copyright 2015, American Institute of Physics. (b) Superhydrophobic-like bouncing phenomenon on liquid thin film at We = 10. Reproduced with permission.[152] Copyright 2015, Nature Publishing Group. (c) Schematic diagram of drop impact on different surfaces with liquid thickness variation ranging from 0 to infinity and the drop bouncing dynamics.

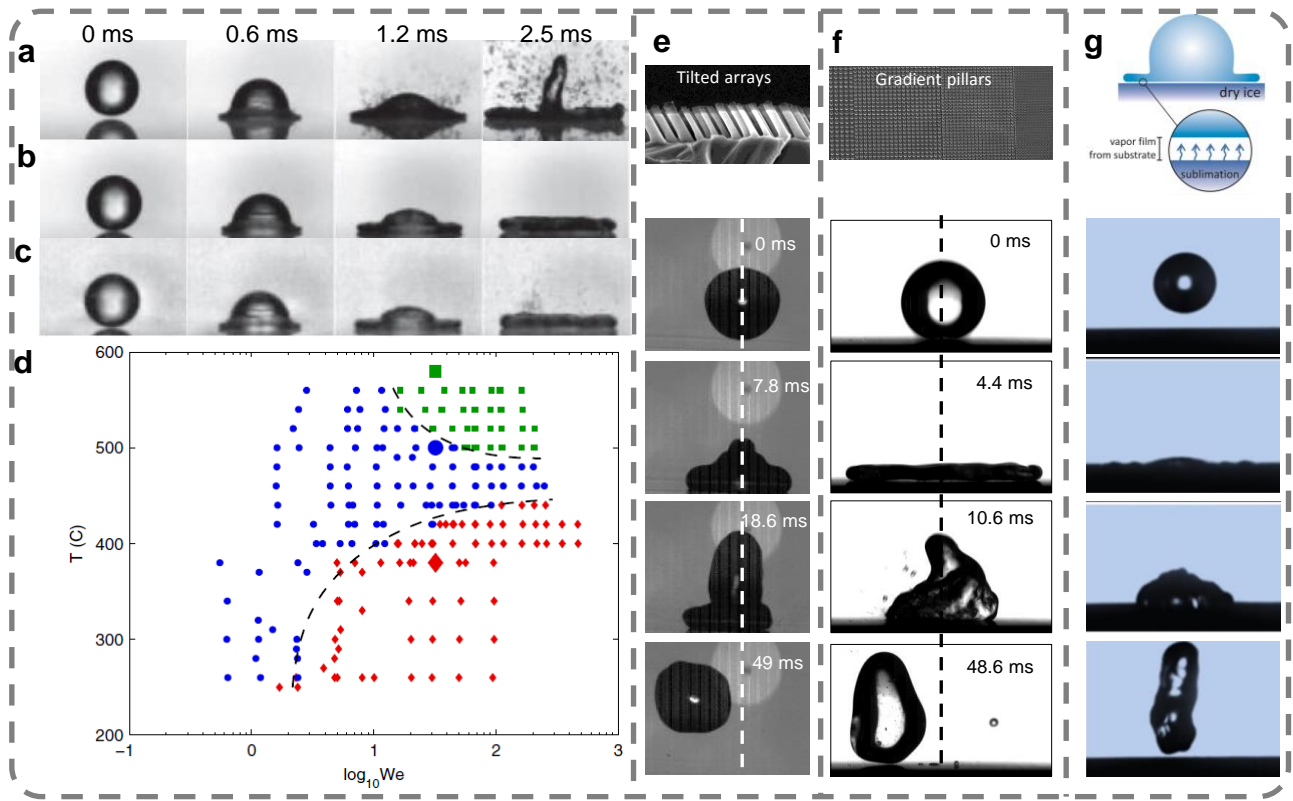

**Figure 6.** Drop bouncing mediated by temperature. (a-c) Representative images of water droplet impacts in contact-boiling regime with temperature $T$ = 380 ℃ (a), Leidenfrost regime with temperature $T$ = 500 ℃ (b), and film boiling regime with temperature $T$ = 580 ℃ (c), respectively. (d) Phase diagram for the impact of water drops on hot plates. For $We > 1$, three phases are observed: a contact-boiling regime (red data) at low temperature, below

the dotted line; a Leidenfrost regime (blue data) at higher temperature; and a spraying film boiling regime (green data) at even higher temperature. Reproduced with permission.[44] Copyright 2012, American Physical Society. (e) Selected snapshots (lower panel) showing the preferential motion of an impinging droplet with *We* = 19.3 on the asymmetric surface (upper panel) at temperature 265 ℃. Image provided by Rebecca Agapov. (f) Selected images (lower panel) demonstrating the directional bouncing of the drop on structural gradient micropillars under high temperature. (g) Image sequence of water drop impacting on solid carbon dioxide at - 79 ℃ (lower panel). Due to the sublimation of dry ice, the vapor film formed and lead to the complete bouncing of droplet (upper panel). Reproduced with permission.[168] Copyright 2013, American Physical Society.

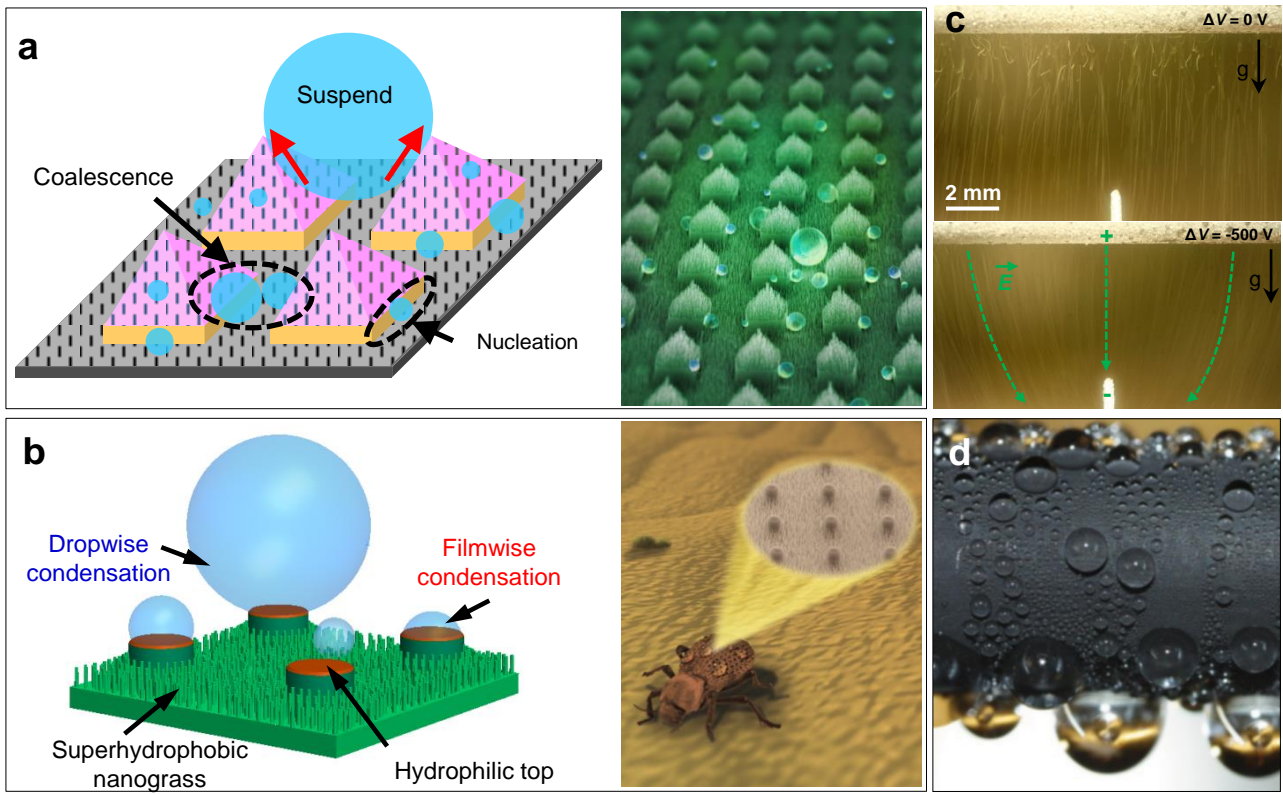

**Figure 7.** Bio-inspired surfaces for enhanced dropwise condensation. (a) Drop condensation on the hierarchical nanograssed micropyramidal architecture (right panel) with the schematic diagram showing that drops preferentially grow on the smooth patches. Reproduced with permission.[114] Copyright 2011, WILEY-VCH. (b) Namib Desert beetle inspired surfaces (right panel) for condensation process with the schematic diagram showing the combination of filmwise and dropwise condensation. The globally dropwise condensation obtained on hybrid surface (inset) demonstrated 63% increase in heat transfer coefficient compared with homogenous hydrophobic surfaces. Reproduced with permission.[115] Copyright 2014, American Chemical Society. (c) images of condensation drops under 0 V bias (upper) with obvious drop-drop interactions and return to the surface against gravity and 500 V bias (lower) with attraction of jumping drops away from the surface. Reproduced with permission.[220] Copyright 2013, American Chemical Society. (d) Image of drop condensation on SLIPS demonstrating higher drop density and low departure radius (0.98 ± 0.13 mm). Reproduced with permission.[224] Copyright 2013, Nature Publishing Group.

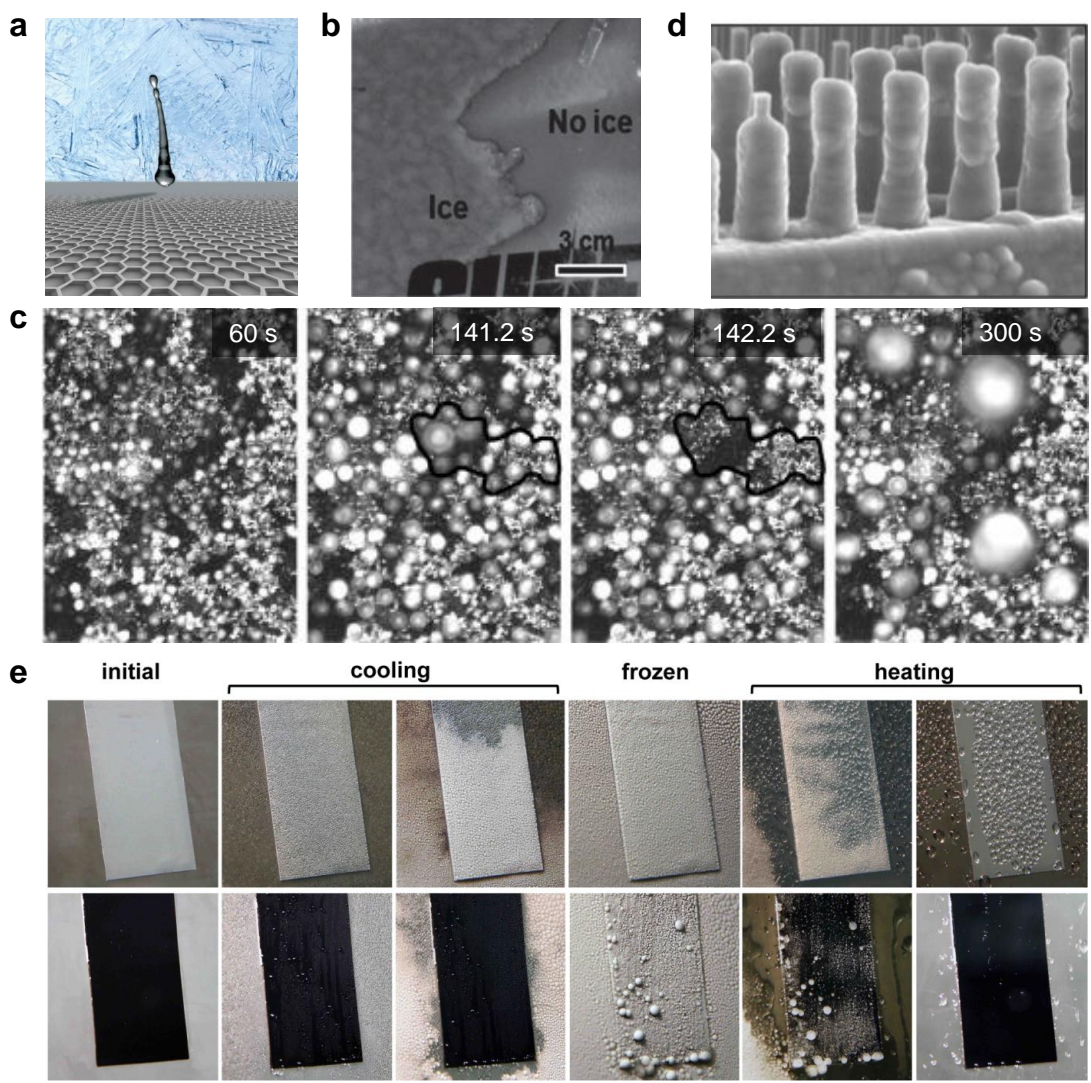

**Figure 8.** Bio-inspired surfaces for anti-icing. (a) Optical image of a closed-cell structure with anti-icing performance during the drop impingement process. Reproduced with permission.[178] Copyright 2010, American Chemical Society. (b) Test of anti-icing properties in naturally occurring "freezing rain" on satellite dish antenna showing that the left side is untreated and is completely covered by ice, while the right side is coated with the superhydrophobic composite and has no ice. Reproduced with permission.[179] Copyright 2009, American Chemical Society. (c) Growth of subcooled condensate on horizontally oriented SHS chilled to –10 °C demonstrating that drops on the SHS were able to continuously jump off the surface before freezing could occur. Reproduced with permission.[234] Copyright 2013, American Chemical Society. (d) Snapshot images of frost formation on the SHS demonstrating the frost nucleation and growth occurs without any particular spatial preference on all of the available area including post tops, sidewalls, and valleys due to the uniform intrinsic wettability of the surface. Reproduced with permission.[242] Copyright 2010, American Institute of Physics. (e) Images showing ice formation by deep freezing (–10 °C) in high-humidity condition and subsequent deicing by heating on bare Al and SLIPS-Al, respectively. The morphology of accumulated ice on SLIPS-Al is significantly different from that on bare Al. Ice still forms mostly around the edges of SLIPS-Al by bridging from the surrounding aluminum substrate (lower panel), while it forms uniformly all over the aluminum substrate (upper panel). Reproduced with permission.[180] Copyright 2013, American Chemical Society.